\documentclass[5p,sort&compress]{elsarticle}

\usepackage[switch]{lineno}

\usepackage{amssymb,amsmath}
\usepackage{lmodern}
\usepackage{bm}
\usepackage{sansmath}
\usepackage{mathrsfs}
\usepackage{url}
\usepackage{upgreek}
\usepackage{mathpazo}
\usepackage{booktabs}

\newcommand{\degree}{\ensuremath{^\circ}}
\newcommand{\lambdaaerzero}{$\Lambda^0_{\rm{aer}}$~}

\newcommand{\haerzero}{$H^{0}_{\rm{aer}}$}

\journal{Astroparticle Physics}

\begin{document}

\begin{frontmatter}


 \author[label1]{Karim Louedec\corref{test}}
 \corref{test}
 \ead{karim.louedec@lpsc.in2p3.fr}
 \author[label1]{Joshua Colombi}
 \address[label1]{Laboratoire de Physique Subatomique et de Cosmologie (LPSC), Universit\'e Grenoble-Alpes, CNRS/IN2P3, 38026 Grenoble cedex, France}

\title{Atmospheric multiple scattering of fluorescence light from extensive air showers and effect of the aerosol size on the reconstruction of energy and depth of maximum}


\begin{abstract}
The reconstruction of the energy and the depth of maximum $X_{\rm max}$ of an extensive air shower depends on the multiple scattering of fluorescence photons in the atmosphere. In this work, we explain how atmospheric aerosols, and especially their size, scatter the fluorescence photons during their propagation. Using a Monte Carlo simulation for the scattering of light, the dependence on the aerosol conditions of the multiple scattered light contribution to the recorded signal is fully parameterised. A clear dependence on the aerosol size is proposed for the first time. Finally, using this new parameterisation, the effect of atmospheric aerosols on the energy and on the $X_{\rm max}$ reconstructions is presented for a vertical extensive air shower observed by a ground-based detector at $30~$km: for typical aerosol conditions, multiple scattering leads to a systematic over-estimation of $5\pm1.5\%$ for the energy and $4.0\pm 1.5~$g/cm$^2$ for the $X_{\rm max}$, where the uncertainties refer to a variation of the aerosol size.
\end{abstract}


\end{frontmatter}



\section{Introduction}
The detection of ultra-high energy cosmic rays using nitrogen fluorescence emission induced by extensive air showers is a well established technique~\cite{sokolsky}, used previously by the Fly's Eye~\cite{flyeye} and HiRes~\cite{hires} experiments, currently by the Pierre Auger Observatory~\cite{pao,pao_fd} and Telescope Array~\cite{ta,ta_fd}, and possibly soon by the JEM-EUSO telescope~\cite{jemeuso}. Charged particles generated during the development of extensive air showers excite atmospheric nitrogen molecules that emit fluorescence light isotropically in the $300-430~$nm range. The energy and geometry of the extensive air shower can then be calculated from information on the amount and time of recorded light signals at the fluorescence detectors (FD). After more than thirty years of development having led to a better understanding of this technique, the current "hybrid" observatories set their energy scale using fluorescence measurements~\cite{ICRC_Verzi,ICRC_Ikeda}. Also, the air-fluorescence technique allows the determination of the depth of maximum of the extensive air shower $X_{\rm max}$ in a direct way, providing an estimation of the UHECR composition~\cite{Auger_Xmax,TA_Xmax}. For the greatest energies, fluorescence light from an air shower can be recorded at distances up to about $40~$km, traversing a large amount of atmosphere before reaching the detector. The effects of the atmosphere on the propagation and attenuation of light must hence be considered carefully. From the production point to the detector, a fluorescence photon can be scattered and/or absorbed by molecules and/or aerosols in the atmosphere. In the case of hazy conditions or fog, the single light scattering approximation -- when scattered light cannot be dispersed again to the detector and only direct light is recorded -- is not valid anymore. Thus, the multiple light scattering -- when photons are scattered several times before being detected -- has to be taken into account in the total signal recorded. Whereas the first phenomenon reduces the amount of fluorescence photons arriving at the telescope, the latter increases the total signal recorded and the apparent angular width of the shower track (i.e.\ the point spread function of the detector). This atmospheric blur occurs especially for long distances to the extensive air shower and total optical depth values greater than unity. Ignoring this contribution to the total light recorded at the fluorescence telescopes would lead to a systematic over-estimation of shower energy and $X_{\rm max}$. In the case of an isotropic point source -- an air shower being usually modelled as a collection of point sources -- , it was highlighted that aerosol scattering is the main contribution to atmospheric blur~\cite{Dave,Pearce,Kopeika_1,Kopeika_2}, resulting especially from aerosol scatter of light at near-forward angles~\cite{Kopeika_3,Reinersman}. Another radiation is also produced in extensive air showers: the Cherenkov light, emitted mostly at small angles relative to the shower axis. This anisotropic emission produces distributions of scattered light different than in the case of fluorescence light. Therefore the work presented in the following can not be applied to the multiple  scattering of Cherenkov light.
~\\

Three main studies about multiple scattering effect on air shower reconstruction have been done during the last ten years: two of them based on Monte Carlo simulations~\cite{MS_roberts,MS_pekala}, and the last one using only analytical calculations~\cite{MS_giller}. Contrary to analytical solutions, Monte Carlo simulations allow to follow each photon or photon packet emitted by an air shower and provide their number of scatterings during the propagation, and their arrival direction and time at the detector. All of these works predict the percentage of indirect light recorded at the detector within its time resolution (usually $100~$ns) and within a circle of angular radius $\zeta$, for every shower geometry and aerosol conditions. Moreover, these parameterisations are currently used in UHECR observatories to remove the multiple scattered fraction from total signal recorded by the fluorescence telescopes. The multiple scattering of light is affected by the optical thickness of the atmosphere, the aerosol size distribution and the aerosol vertical profile. Whereas the three previous works have studied the effect of the optical thickness, the multiple scattering is also very dependent on the aerosol size distribution, and especially on the corresponding asymmetry parameter of the aerosol scattering phase function~\cite{MyThese}. This dependence was recently studied in detail in~\cite{MS_louedec_PS,MS_pekala_PS} for the case of an isotropic point source. The purpose of this paper is to quantify the dependence of the percentage of indirect light on the aerosol size, and its corresponding effect on the shower energy and $X_{\rm max}$ reconstructions. Section~\ref{sec:modelling} is a brief introduction of some quantities concerning light scattering, before describing in detail the Monte Carlo simulation developed and adapted for this work. Section~\ref{sec:simu_results} gives a general overview of properties of scattered photons recorded at the detector for different atmospheric conditions. Then, in Section~\ref{sec:parameterisation}, we propose an improved parameterisation based on the previous work done in~\cite{MS_roberts}, but including this time an explicit dependence on the aerosol size. This result is finally applied to the reconstruction of extensive air showers in the case of a ground-based detector in Section~\ref{sec:eas}: the effect on energy and $X_{\rm max}$ reconstructions is given for different aerosol conditions and distances to the air shower.

\section{Modelling and simulation of scattering in the atmosphere}
\label{sec:modelling}
Throughout this paper, the scatterers in the atmosphere will be modelled as non-absorbing spherical particles of different sizes~\cite{Hulst,Bohren}. It is acceptable to approximate the scatterers as being non-absorbing, since absorption (mainly by ozone and dioxide nitrogen) is negligible compared to scattering in the atmosphere in the range of wavelengths $300-430~$nm of fluorescent UV light. Scatterers in the atmosphere are usually divided into two main types -- aerosols and molecules.

\subsection{The density of scatterers in the atmosphere}
The attenuation length (or mean free path) $\Lambda$ associated with a given scatterer is related to its density and is the average distance that a photon travels before being scattered. For a given number of photons $N$ traveling across an infinitesimal distance d$l$, the amount scattered is given by ${\rm d}N^{\rm{scat}} = N \times
{\rm d} l / \Lambda$. Density and $\Lambda$ are inversely related such that a higher value of $\Lambda$ is equivalent to a lower density of scatterers in the atmosphere. Molecules and aerosols have different associated densities in the atmosphere and are described respectively by a total attenuation length $\Lambda_{\rm{mol}}$ and $\Lambda_{\rm{aer}}$.  The value of these total attenuation lengths in the atmosphere can be modelled as horizontally uniform and exponentially increasing with respect to height above ground level $h_{\rm agl}$. The total attenuation length for each scatterer population is written as 
\begin{equation}
\begin{cases}
\Lambda_{\rm mol}(h_{\rm agl}) = \Lambda^{0}_{\rm mol} \,\exp \left[ (h_{\rm agl} + h_{\rm det})/H^{0}_{\rm mol} \right],\\
\Lambda_{\rm aer}(h_{\rm agl}) = \Lambda^{0}_{\rm aer} \, \exp\left[ h_{\rm agl}/ H^{0}_{\rm aer} \right],
\end{cases}
\label{eq:lambdas}
\end{equation}
where $\{\Lambda^{0}_{\rm{aer}}$, $\Lambda^{0}_{\rm{mol}}\}$ are multiplicative scale factors, $\{H^{0}_{\rm{aer}}$, $H^{0}_{\rm{mol}}\}$ are scale heights associated with aerosols and molecules, respectively, and $h_{\rm det}$ is the altitude difference between ground level and sea level (fixed at $1400~$m for the whole study). The \emph{US standard atmospheric model} is used to fix typical values for molecular component: $\Lambda^{0}_{\rm{mol}} = 14.2~$km and $H^{0}_{\rm{mol}} = 8.0~$km~\cite{Bucholtz}. These values are of course slightly variable with weather conditions~\cite{EPJP_BiancaMartin} but the effect of molecule concentration on multiply scattered light is not that of interest in this work. Atmospheric aerosols are found in lower densities than molecules in the atmosphere and are mostly present only in the first few kilometres above ground level. The aerosol population is much more variable in time than the molecular as their presence is dependent on many more factors such as the wind, rain and pollution~\cite{EPJP_Aerosol}. However, the model of the exponential distribution is usually used to describe aerosol populations~\cite{pao_vaod}. Only the parameter \lambdaaerzero will be varied and \haerzero~is fixed at $1.5~$km for the entirety of this work.

\subsection{The different scattering phase functions}
\label{sec:phase_functions}
A scattering phase function is used to describe the angular distribution of scattered photons. It is typically written as a normalised probability density function expressed in units of probability per unit of solid angle. When integrated over a given solid angle $\Omega$, a scattering phase function gives the probability of a photon being scattered with a direction that is within this solid angle range. Since scattering is always uniform in azimuthal angle $\phi$ for both aerosols and molecules, the scattering phase function is always written simply as a function of polar scattering angle $\psi$.\\

Molecules are governed by Rayleigh scattering which can be derived analytically via the approximation that the electromagnetic field of incident light is constant across the small size of the particle~\cite{Bucholtz}. The molecular phase function is written as
\begin{equation}
P_{\rm{mol}}(\psi)=\frac{3}{16 \pi}(1+\cos^2\psi),
\label{eq:RPF}
\end{equation}
where $\psi$ is the polar scattering angle and $P_{\rm{mol}}$ the probability per unit solid angle. The function $P_{\rm{mol}}$ is symmetric about the point $\pi/2$ and so the probability of a photon scattering in forward or backward directions is always equal for molecules.

\begin{figure}[t!]
\centering
\vspace{-0.8cm}
\includegraphics[width = 0.5\textwidth]{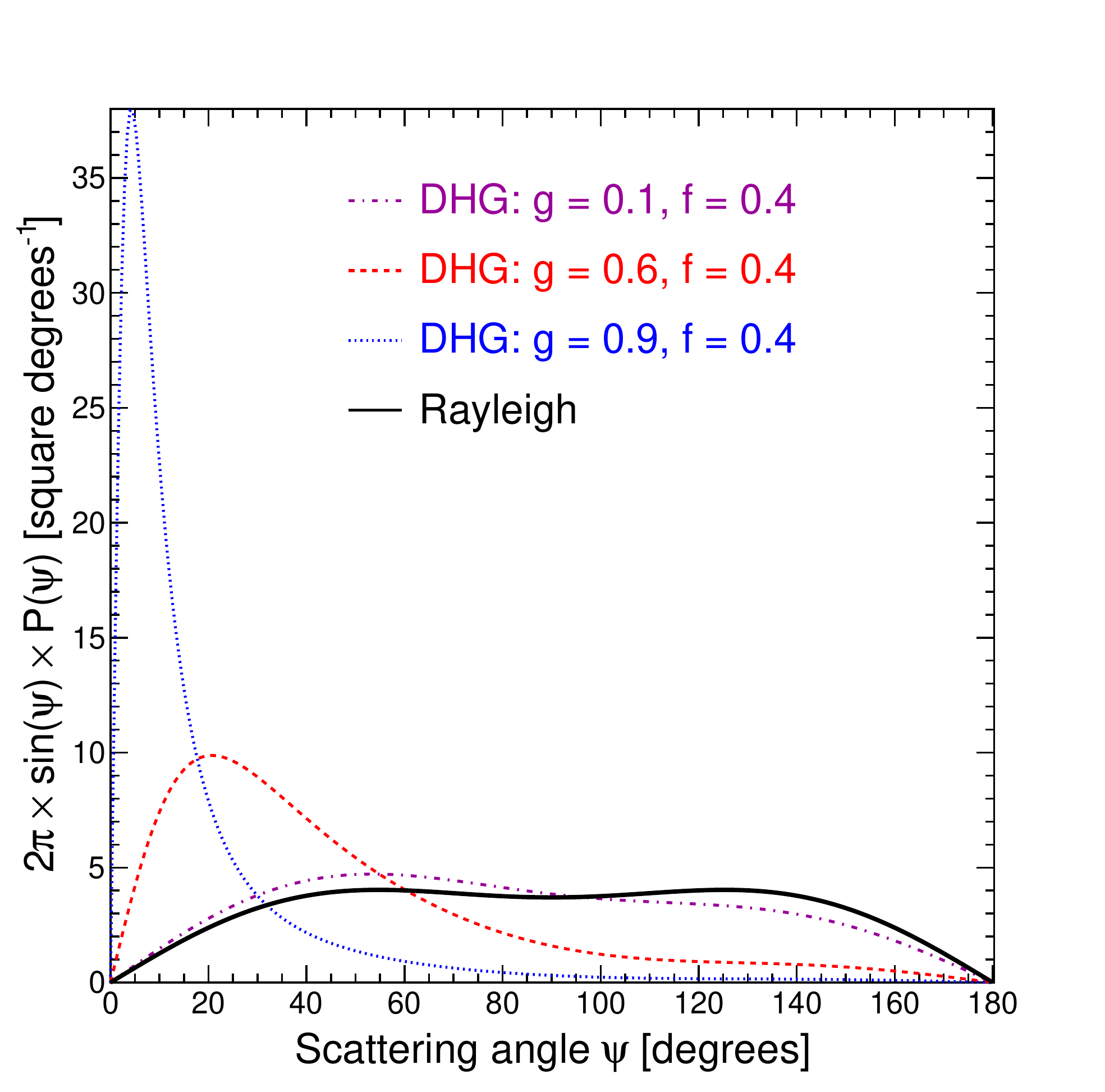}
\caption{{\bf Scattering phase function per unit of polar angle $\psi$, and its dependence to the asymmetry parameter.} Scattering phase functions are in units of probability per solid angle $\Omega$ as opposed to probability per unit of $\psi$ as necessary to get the probability density function of the polar angle $\psi$. Thus, the scattering phase functions $P_{\rm{mol}} (\psi)$ and $P_{\rm{aer}} (\psi)$ have to be multiplied by $2\pi \,\sin{\psi}$ to remove the solid angle weighting. Colours in online version.}
\label{fig:APF}
\end{figure}

Atmospheric aerosols typically come in the form of small particles of dust or droplets found in suspension in the atmosphere. The angular dependence of scattering by these particles is less easily described as the electromagnetic field of incident light can no longer be approximated as constant over the volume of the particle. Mie scattering theory~\cite{Mie} offers a solution in the form of an infinite series for the scattering of non-absorbing spherical objects of any size. Since this infinite series is far too time consuming for the Monte Carlo simulations, a parameterisation named the Double Henyey-Greenstein (DHG) phase function~\cite{HenyeyGreenstein,EPJP_Aerosol} is usually used. It is a parameterisation valid for various particle types and different media~\cite{HG_astro,HG_meteo,HG_bio}. It is written as
\begin{eqnarray}
P_{\rm{aer}}(\psi|g,f)= &\frac{1-g^2}{4\pi}\left[\frac{1}{(1+g^2-2g\cos{\psi})^\frac{3}{2}} +f\left(\frac{3\cos^2{\psi}-1}{2(1+g^2)^\frac{3}{2}}\right)\right]
\label{eq:APF}
\end{eqnarray}
where $g$ is the asymmetry parameter given by $\left<\cos\psi\right>$ and $f$ the backward scattering correction parameter. $g$ and $f$ vary in the intervals $[-1,1]$ and $[0 ,1]$, respectively. Most of the atmospheric conditions can be probed by varying the value of the asymmetry parameter $g$: aerosols ($0.2 \leq g \leq 0.7$), haze ($0.7 \leq g \leq 0.8$), mist ($0.8 \leq g \leq 0.85$), fog ($0.85 \leq g \leq 0.9$) or rain ($0.9 \leq g \leq 1.0$)~\cite{Metari}. The DHG function reduces to the molecular scattering phase function for $\{g=0.0;\,f=0.5 \}$. Changing $g$ from 0.2 to 1.0 increases greatly the probability of scattering in the very forward direction as it can be observed in Fig.~\ref{fig:APF}. The reader is referred to ~\cite{Ramsauer_1,Ramsauer_2} to see the recently published work on the relation between $g$ and the mean radius of an aerosol. A physical interpretation of the asymmetry parameter $g$ in the DHG phase function is the mean aerosol size: the amount of light scattered in the forward direction is proportional to the square of the aerosol radius and the width of the forward-scattering peak is proportional to the inverse of the aerosol radius. Measurements of the aerosol phase function were operated by the HiRes~\cite{APF_hires} and the Pierre Auger~\cite{APF_pao,MyICRC} collaborations: a typical value found for the asymmetry parameter $g$ is 0.6. In the case of an incident light with a wavelength of $400~$nm, a $g$ value of $0.6$ corresponds to a mean aerosol radius of about $0.2~\upmu$m~\cite{Ramsauer_2}. This measured value does not take into account a possible evolution of the aerosol size with the altitude, but it can be considered as an effective $g$ value for the lowest part of the atmosphere. Since fluorescence detectors do not operate during foggy or rainy conditions, only cases with $g$ values lower than $0.8$ will be studied in the following. The parameter $f$ is an extra parameter acting as a fine tune for the amount of backward scattering. It will be fixed at 0.4 for the rest of this work as this is the value typically measured in desert like conditions.

\subsection{Monte Carlo code description}
\label{sec:code}
Multiple scattering of fluorescence photons was simulated using a Monte Carlo code developed originally to study multiple scattering from an isotropic point source to a detector~\cite{MS_louedec_PS}. This Monte Carlo simulation traces individually every multiply scattered photon detected at the fluorescence detectors. An isotropic light source is simulated by creating $N$ photons with the same initial position and isotropically distributed initial directions. Photons are then propagated through a given distance $D$ corresponding to the distance between the light source and the ground-based detector. A photon is randomly scattered by an aerosol, molecule or not at all in accordance with the probabilities ${\rm d}l/{\Lambda_{\rm{aer}}}$, ${\rm d}l/{\Lambda_{\rm{mol}}}$ or $1 - {\rm d}l/{\Lambda_{\rm{aer}}} - {\rm d}l/{\Lambda_{\rm{mol}}}$, respectively. The scattering angle is dependent on the scattering phase function involved: the Rayleigh and the Double Henyey-Greenstein scattering phase functions are used for molecular and aerosol scattering events, respectively. A horizontally uniform density distribution for aerosols and molecules is assumed for the vertical profile of the atmosphere. The present work does not investigate the effect of a change of the vertical distribution of aerosols (i.e.\ exponential shape and vertical aerosol scale $H^{0}_{\rm{aer}}$), nor the effect of overlying cirrus clouds or aerosol layers on the multiple scattered light contribution to direct light.

\begin{figure*}[t!]
\centering
\includegraphics[width = 0.48\textwidth]{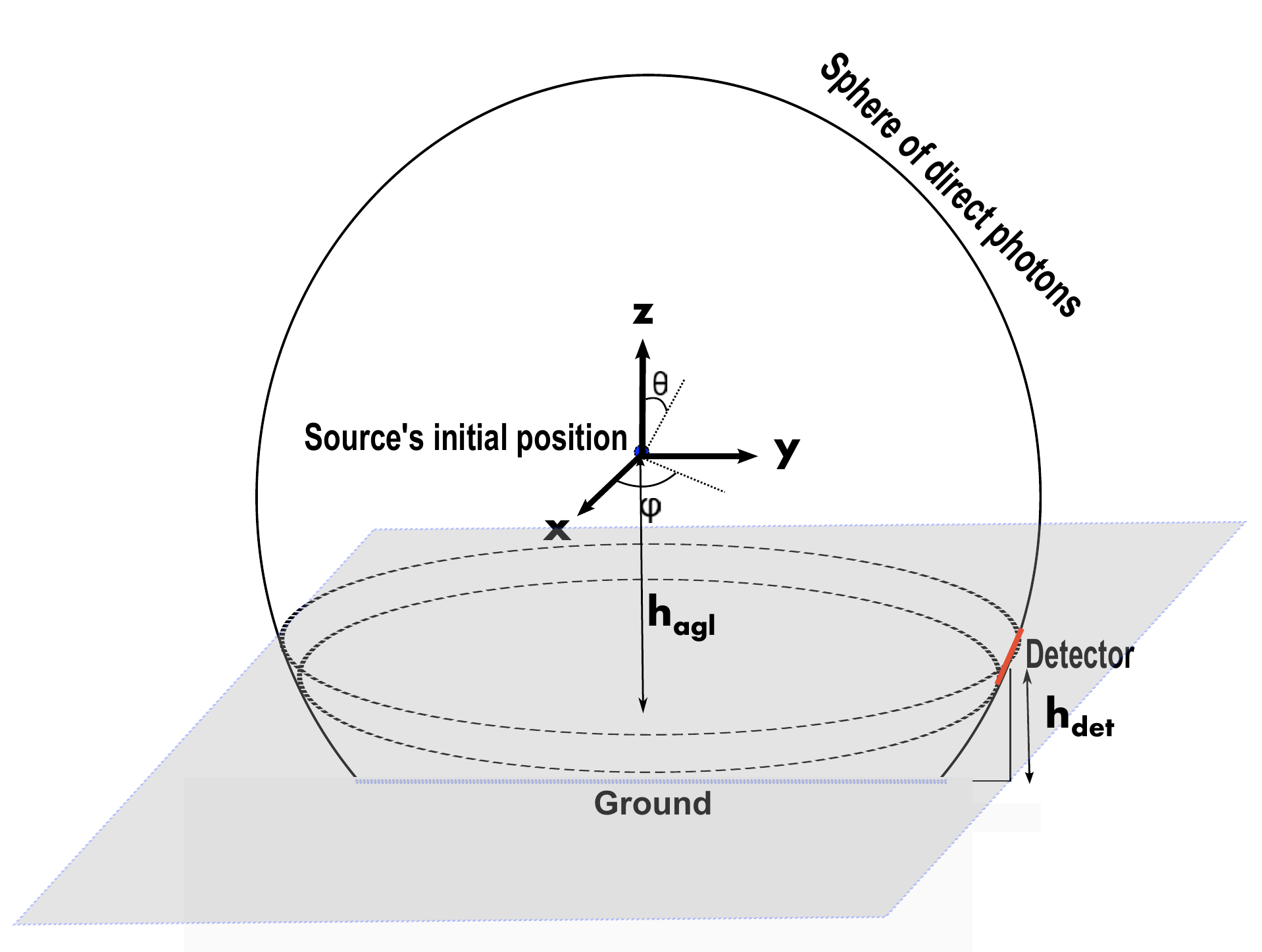}
\includegraphics [width=0.50\textwidth] {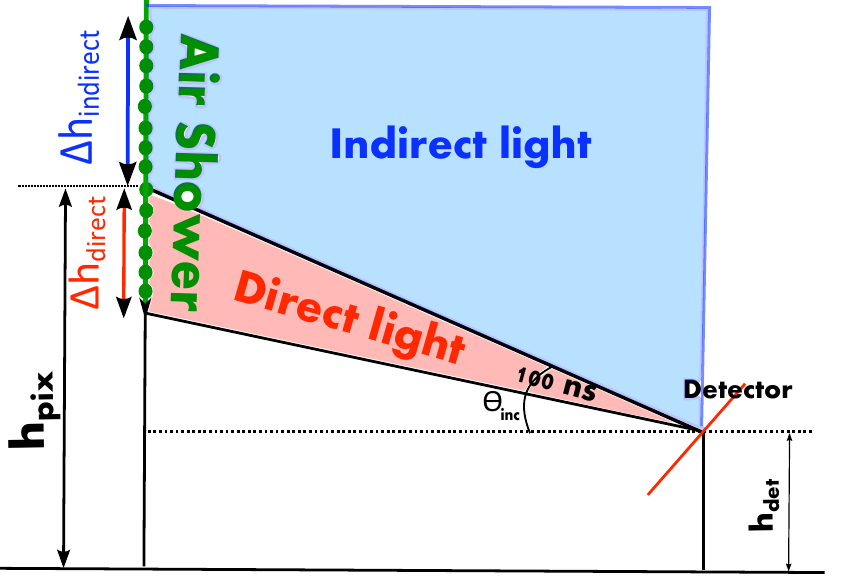}
\caption{{\bf (left) Sketch of an isotropic point source and a ground-based detector.} A diagram showing how the detector is simulated to have an extent of $2\pi$ in azimuthal angle to increase the amount of statistics retrieved for indirect photons. {\bf (right) Geometrical setup of a vertical air shower (i.e.\ a cascade of isotropic sources) in the simulation at a core distance $L$ and an inclination angle $\theta_{\rm inc}$.} The pixel is activated at the time that direct light from $h_{\rm pix}$ arrives at the detector and then remains receptive to light for a time period of $100~$ns later. Regions of space contributing to direct light and indirect light are given in red and blue, respectively. Colours in online version.}
\label{fig:sketch}
\end{figure*}

The quantity of multiple scattered light recorded by a fluorescence telescope is of principal interest, and especially this contribution as a function of the integration angle $\zeta$. The angle $\zeta$ is defined as the angular deviation in the entry of indirect photons at the detector aperture with respect to direct photons. For direct photons from an isotropic source, the angle of entry is usually approximated to be constant as the entry aperture of the detector is always very small relative to the distance of the isotropic sources. In contrast, multiply scattered photons can enter the aperture of the detector at any deviated angle $\zeta$ from the direct light between $0^\circ$~and $90^\circ$. The main problem in simulating indirect light contribution at detectors is obtaining reasonable statistics within reasonable simulation running times. The root of the problem is the very small surface area of the detector relative to the large distances where fluorescence photons are created. The amount of direct photons is calculated analytically by modelling the detector as a point relative to the initial position of the isotropic source. The same approach as explained in~\cite{MS_roberts} is used to cut running times of the simulation for indirect photons. The symmetry in the distribution of scatterers in azimuthal angle -- due to the horizontal uniformity of the scatterers --  is applied here. This symmetry means that so long as the detector has the same height and distance from the source, the azimuthal angle relative to it is unimportant. As such, the surface area of the detector is increased in the simulation by extending it through an azimuthal angle of $2\pi$ so that a greater amount of indirect photons is detected and better statistics are obtained. The setup of the extended detector is drawn in Fig.~\ref{fig:sketch}~(left), where the strip of the sphere has a width corresponding to the diameter of the detector. Also, stopping the tracking of all photons that can no longer be detected further reduces the simulation time.

In order to simulate multiple scattering of fluorescence light emitted by an extensive air shower, modifications to the original program were made. It is well-known that fluorescence light is emitted isotropically in space, thus an air shower can be approximated to a cascade of isotropic sources separated in time and space by the speed of light $c$.  The study is carried only for vertical air showers -- i.e.\ a vertical column of isotropic sources --, since results in~\cite{MS_roberts} show that it is a good approximation for air showers of any geometries too (this assumption based on the spatial origin of the scattered photons will be checked later in Sect.~\ref{sec:simu_results}). A typical fluorescence detector use large area mirrors to focus the light emitted by extensive air showers onto arrays of photomultiplier tubes (PMT). A single PMT -- corresponding to one pixel -- has a field-of-view in elevation and azimuth of $\sim 1^\circ$ in the case of Telescope Array and $1.5^\circ$ in the case of the Pierre Auger Observatory. Thus, such pixels can observe the propagating front of an air shower for periods of time up to several hundreds of nanoseconds. However, the read-out time associated with the FDs is usually 100~ns. It is the contribution of indirect light relative to direct light for these given $100~$ns time frames that is of interest. In the simulation of air showers, a specific inclination angle $\theta_{\rm inc}$ is used to describe the point at which direct light from the air shower first enters the fluorescence detector and the time frame of $100~$ns begins. A core distance $L$ is used to denote the distance along the ground from the detector to the vertical air shower. Light arriving from all isotropic sources within the given $100~$ns time frame at the detector is accounted for. The geometry of the situation is set out in Fig.~\ref{fig:sketch}~(right). The complete simulation of an air shower would begin at the edge of the atmosphere ($\sim20$~km above ground level) and means its simulation is too long, even for modern day computer resources available. The solution is to find a balance between a range of heights giving a good approximation of the extent of the air shower whilst equally allowing a sufficient amount of statistics on indirect photons to be collected. There is a finite space across which isotropic sources in the air shower can contribute to the direct light signal recorded at the detector. The extent of this height is denoted $\Delta h_{\rm direct}$ and is given by the equation
\begin{equation}
\Delta h_{\rm direct} = \frac{ \gamma^2 - L^2 + h_{\rm{pix}} - h_{\rm{det}} } {2( \gamma + h_{\rm{det}} - h_{\rm{pix}} )},
\end{equation}
where $h_{\rm pix}$ is the first height above ground level at which direct light from the air shower is detected and $L$ is the core distance of the shower from the FD. $\gamma$ is equal to $c t_{\rm{det}}  + \sqrt{L^2 + (h_{\rm{pix}} - h_{\rm{det}})^2 }$, with  $t_{\rm det}$ the time window for collecting light at the FD (100~ns here). These equations can be found by considering the geometry shown in Fig.~\ref{fig:sketch}~(right). Then, the direct light signal recorded by the detector can be calculated analytically considering a value of $h_{\rm source}$ approximated by $h_{\rm{pix}} - \Delta h_{\rm{direct}}/2$. The indirect light contribution cannot be calculated analytically and is retrieved from the simulation. In contrast to the direct light, the indirect light arriving at the detector within the 100~ns time frame can originate from sources at an infinite range of heights higher than $h_{\rm{pix}} - \Delta h_{\rm{direct}}/2$, i.e.\ at any time after the detection of direct light. What is required is the height at which indirect sources above $h_{\rm{pix}}$ begin to have a negligible effect on the amount of light seen in the $100~$ns time frame. This range of heights is denoted as $\Delta h_{\rm{indirect}}$. Its value $\Delta h_{\rm{indirect}}$ is more neatly expressed as a time $t_{\rm approx}$, defined as the time after direct light from a source has been detected that indirect light from the same source can be approximated to no longer have an effect. The equation relating $\Delta h_{\rm{indirect}}$ and $t_{\rm approx}$ can be derived as 
\begin{equation}
\Delta h_{\rm{indirect}} = \frac{ L^2 + h_{\rm{pix}} - h_{\rm{det}} -\beta^2 } {2( \beta + h_{\rm{det}} - h_{\rm{pix}} )},
\end{equation}
where $\beta$ is given by $\sqrt{L^2 + (h_{\rm{pix}} - h_{\rm{det}})^2 } - c t_{\rm{approx}}$.

\begin{figure}[!t]
\centering
\vspace{-0.6cm}
\includegraphics [width=0.50\textwidth] {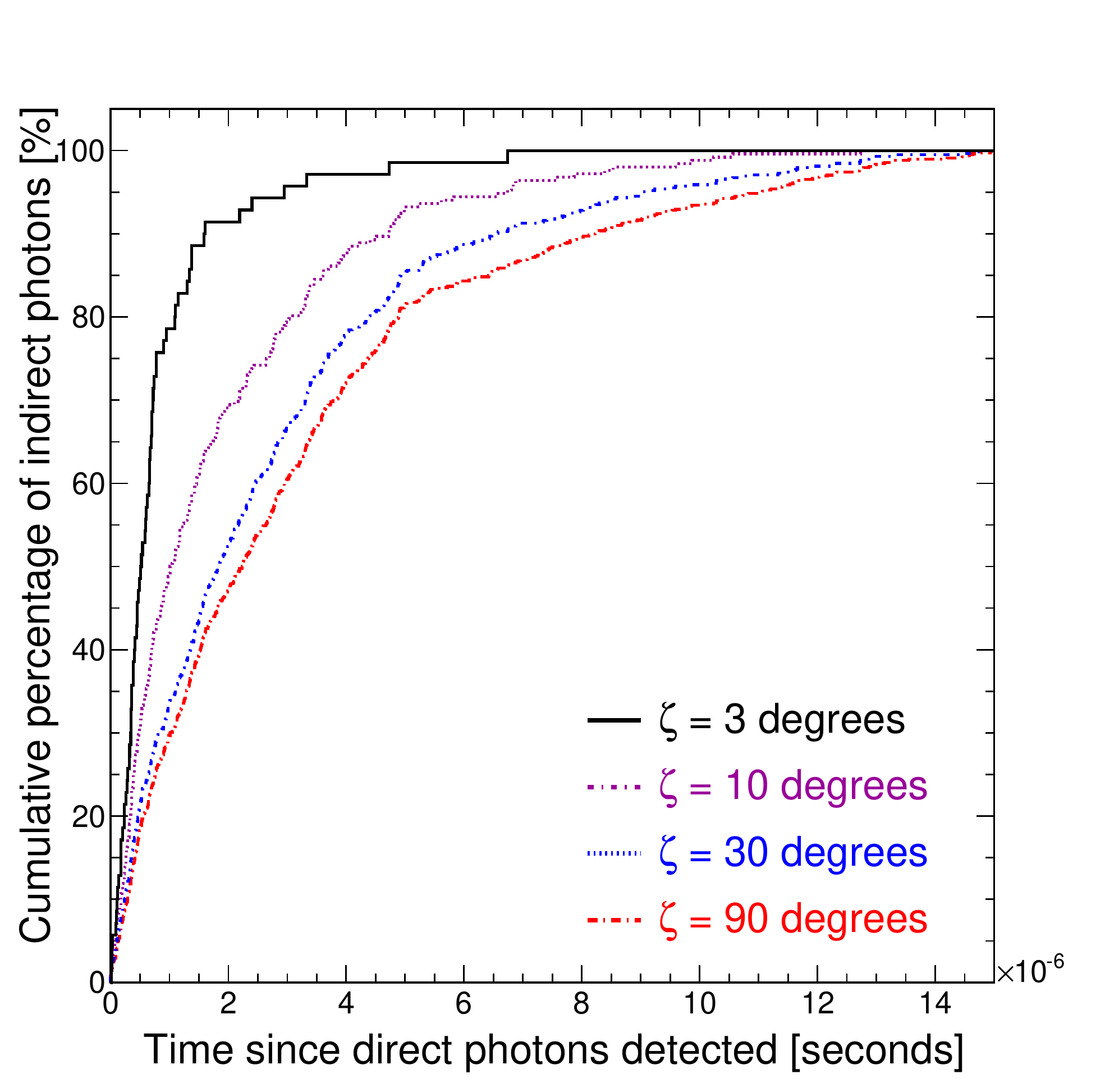}
\caption{{\bf Cumulative percentage of indirect photons originating from isotropic sources whose light at the detector was detected a given of time ago.} The percentage is measured relative to a maximum time $t_{\rm approx}$ of $15000~$ns, after which, indirect photons from an isotropic source are assumed to be no longer detectable. The shower is located $15~$km from the detector and observed with an inclination angle of $15^\circ$. The aerosol attenuation length \lambdaaerzero is fixed at $25~$km and the $g$ parameter is equal to $0.3$. Colours in online version.}
\label{fig:cumul_time}
\end{figure}

In order to fix the value of $t_{\rm{approx}}$ needed to reproduce properly the indirect light from an air shower, simulations are run with a greater than necessary $t_{\rm{approx}}$ of $15000~$ns. The cumulative contribution of indirect light from isotropic sources with direct light detected up to this amount of time ago is displayed in Fig.~\ref{fig:cumul_time}. The results shown are for a configuration with an aerosol attenuation length at ground level of \lambdaaerzero= 25~km, a core distance $L=15$~km, an inclination $\theta_{\rm{inc}}=15\degree$ and an asymmetry parameter $g=0.3$. A lower value of $t_{\rm{approx}}$ at which sources become negligible for indirect light recorded at the detector can be found for each integration angle $\zeta$. For a small integration angle, here $\zeta = 3^\circ$, all indirect light coming later than $\sim5000$~ns is negligible, with only $\sim3\%$ of the total amount of indirect photons arriving after this time. These results and the trend observed for greater $\zeta$ values were expected since light recorded with a longer delay corresponds to indirect photons arriving with a larger variation in entry angle compared to direct light. The probability that the indirect light from a source simulated a long time ago (i.e.\ at a higher height) and being detected within a small integration angle such as $\zeta=3^\circ$ is therefore much less than for a larger integration angle as $\zeta=90^\circ$. The effect of the indirect light from sources at different heights of course varies with aerosol conditions and only one variation is given in Fig.~\ref{fig:cumul_time}. It is noted, however, that a small asymmetry parameter $g=0.3$ is the case where the greatest value of $t_{\rm{approx}}$ is found as indirect light from a greater range of sources along the air shower affect the signal. This will be verified in the next section. Thus, in the rest of this work, all simulations of air showers are run for a value of $t_{\rm{approx}} = 5000$~ns which is a good approximation since the typical value of $\zeta$ for a ground-based cosmic ray detector is $1^\circ$ to $1.5^\circ$. The next section presents a general overview of properties of scattered photons recorded at a ground-based detector for different aerosol conditions.

\section{Properties of indirect photons and multiple scattering contribution fraction}
\label{sec:simu_results}
\begin{figure}[!t]
\vspace{-0.8cm}
\includegraphics [width=0.50\textwidth] {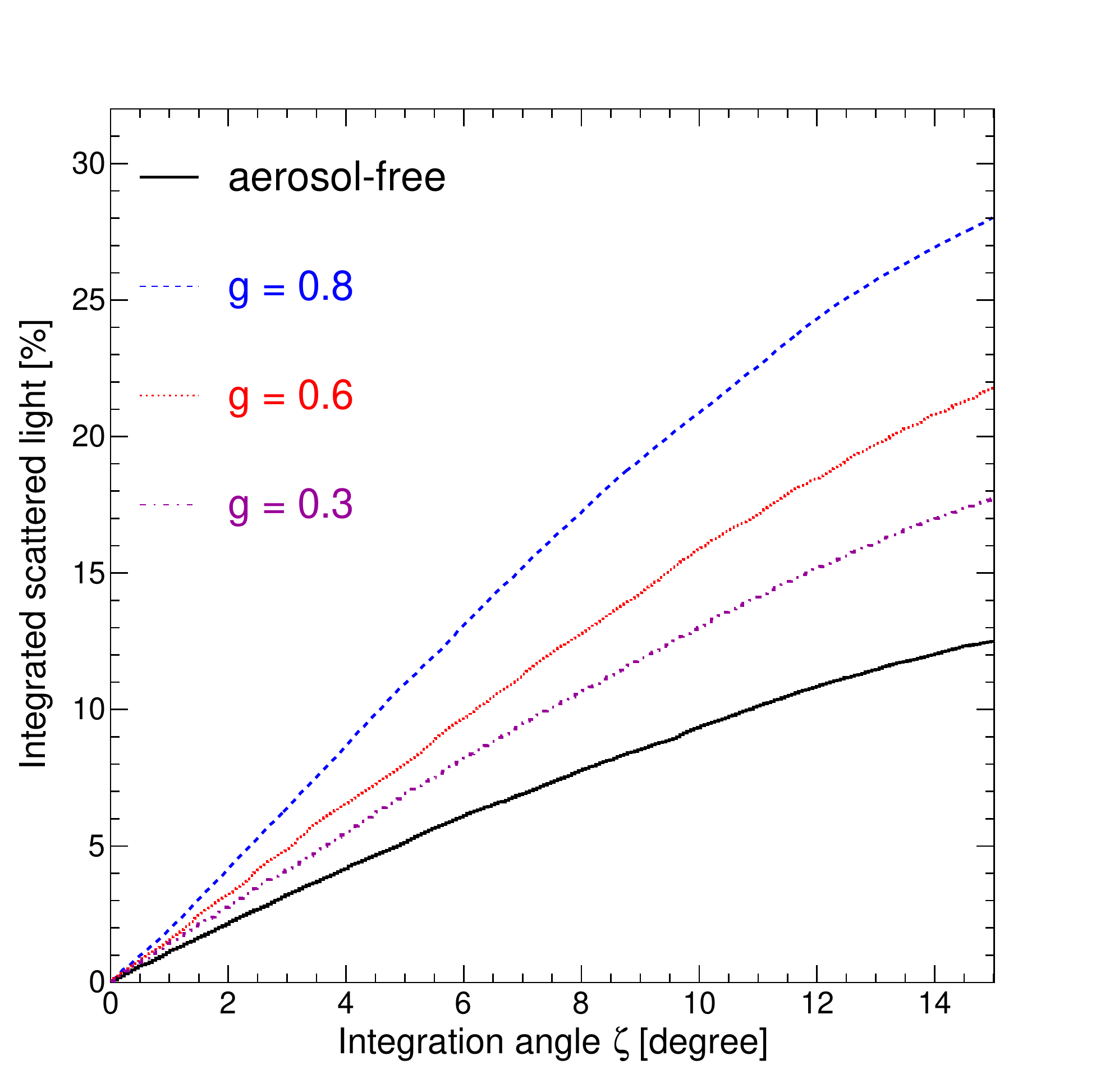}
\caption{{\bf Evolution of the multiple scattering contribution fraction with the integration angle, for different values of asymmetry parameter}. Simulations are done for a vertical shower located at $10~$km from the detector, and it is viewed with a typical inclination angle $\theta_{\rm inc}$ of $12^\circ$. Aerosol conditions are characterised by an aerosol attenuation length equal to $25~$km and an asymmetry parameter $g$ varying from 0.3 to 0.8. The case with an aerosol-free atmosphere is also plotted. Curves represent all indirect photons arriving at the detector during the first $100~$ns. Colours in online version.}
\label{fig:ms_cumulative}
\end{figure}

This part aims to look at the effect of changing aerosol size (via the asymmetry parameter $g$) on the amount of indirect light recorded at the detector for a vertical air shower. The integration time of the detector is set to $100$~ns and the aerosol attenuation length at ground level is fixed at \lambdaaerzero=25~km. The percentage of light due to indirect photons against integration angle $\zeta$ is given by the ratio (indirect light) / (direct light + indirect light), where the direct or indirect light signals are the number of photons collected within the given integration angle $\zeta$. Figure~\ref{fig:ms_cumulative} shows the results for a vertical air shower placed at a core distance $L=10$~km and viewed at an inclination angle of $\theta_{\rm{inc}}= 12^\circ$. As expected, the amount of signal due to indirect light increases consistently with integration angle $\zeta$ as all direct light arrived at $\zeta=0^\circ$. These curves are directly linked to the point spread function since only a differentiation with respect to $\zeta$ is needed. A more interesting feature is the increased contribution from indirect light for increasing aerosol size (i.e.\ a higher value of the asymmetry parameter $g$). It is clearly seen that atmospheric aerosols have a non-negligible effect on the percentage of indirect light received at a ground-based detector, even at low integration angles $\zeta$.

\begin{figure*}[!t]
\centering
\includegraphics [width=1.0\textwidth] {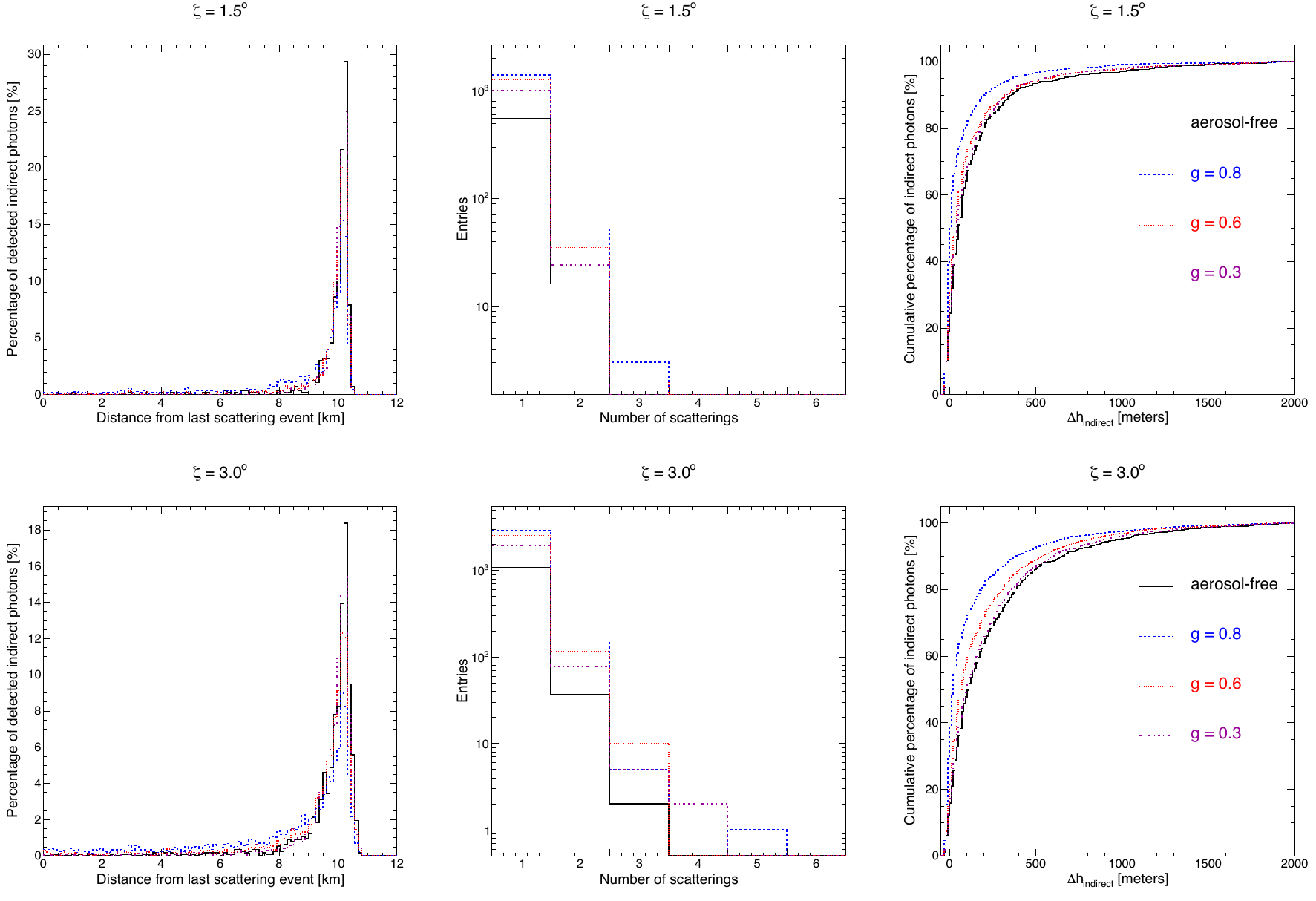}
\caption{{\bf Properties of indirect photons recorded at the ground-based detector and their dependence on the asymmetry parameter}. Simulations are done for a vertical shower located at $10~$km from the detector, and it is viewed with a typical inclination angle $\theta_{\rm inc}$ of $12^\circ$. Aerosol conditions are characterised by an aerosol attenuation length equal to $25~$km and an asymmetry parameter $g$ varying from 0.3 to 0.8. The case with an aerosol-free atmosphere is also given. Plots represent all indirect photons arriving at the detector during the first $100~$ns following the detection of direct light, and an integration angle $\zeta$ of $1.5^\circ$ (top) or $3.0^\circ$ (bottom). Colours in online version.}
\label{fig:ms_properties}
\end{figure*}

In the rest of this section, we will analyse the properties of indirect photons recorded by the ground-based detector in the first $100~$ns following the detection of direct photons, and for an integration angle $\zeta$ of $1.5^\circ$ -- typical value -- or $3.0^\circ$ -- extreme case for light collection. The same vertical shower is still used here to understand multiple scattering of fluorescence photons in the atmosphere. Figure~\ref{fig:ms_properties}~(left) gives the percentage of indirect photons last scattered at a given distance relative to the detector. The difference of shapes between each asymmetry parameter is directly linked to the anisotropy associated with the scattering phase function for atmospheric aerosols. In the case of an atmosphere with very small aerosols tending to molecular size (i.e.\ $g=0.3$ in our case), most of scattered photons recorded within the integration time of $100~$ns originate from the position of photon production. Indeed, there is an increased density of photons at this distance where fluorescence photons are produced. For the case of larger aerosol sizes, corresponding to larger $g$ values, the peak at the position of photon production is now greatly diminished. This is a result of detection of photons scattered more uniformly across all distances. This occurs because aerosols with higher values of $g$ have a higher probability of scattering photons in a very forward direction. Furthermore, with an integration time of $100~$ns as used here, it is expected that most of the multiply scattered photons do not undergo more than one scattering before being detected. It is exactly what we observe in Fig.~\ref{fig:ms_properties}~(middle).

We explained in Sect.~\ref{sec:code} that, even if only vertical air showers are simulated to study multiple scattering of light in the atmosphere in this work, the parameterisation of multiply scattered fraction produced would be valid also for non-vertical air showers. The same approximation is used by M.\ D.\ Roberts in~\cite{MS_roberts}. Figure~\ref{fig:ms_properties}~(right) gives the cumulative percentage of indirect photons as a function of the track length $\Delta h_{\rm direct}$. A value $\Delta h_{\rm direct} = 0$ would indicate that indirect photons come from the same position as direct photons $h_{\rm source}$. Thus, in the usual case where $\zeta = 1.5^\circ$ and the asymmetry parameter $g\geqslant 0.6$, more than $80\%$ of indirect photons come from a region being within $\sim 200~$m of the position of direct light production $h_{\rm source}$. Thus, even if the inclination of the shower track is changed, the geometry of the track segment near $h_{\rm source}$ with respect to the fluorescence detector, and the aerosol conditions in the surroundings of $h_{\rm source}$ will not change drastically. Consequently, if the shower axis is not pointing too closely to the detector, all the results given in this paper can be applied to any geometries of air showers, including the parameterisation of the multiply scattered fraction which will be obtained in Sect.~\ref{sec:parameterisation}.

\begin{figure*}[p]
\centering
\includegraphics [width=1.0\textwidth] {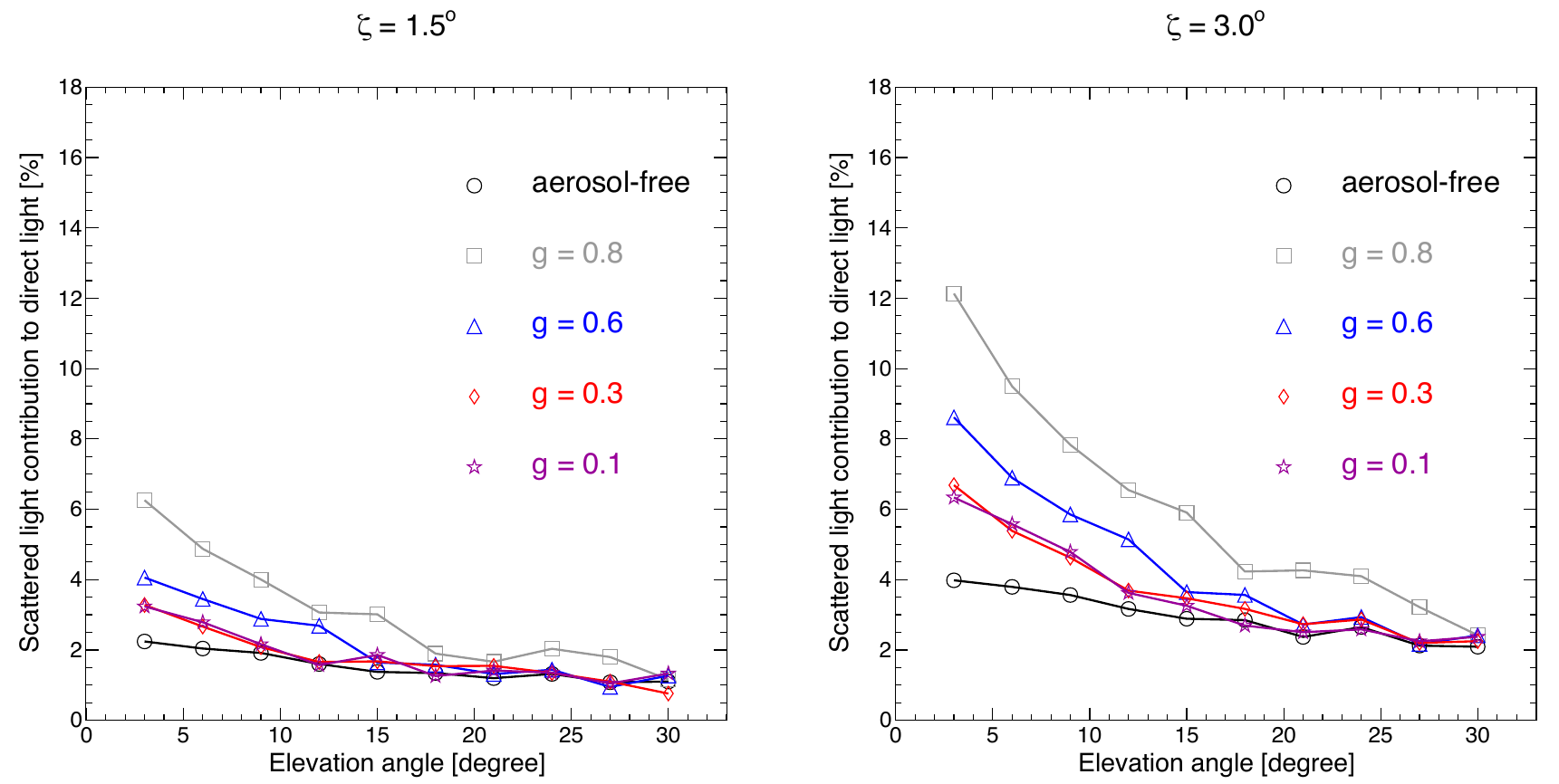}
\caption{{\bf Contribution of the multiple scattered light to the direct light recorded versus the inclination angle of the fluorescence detector, for different values of asymmetry parameter.} Simulations are done for an aerosol attenuation length equal to $25~$km. The case with an aerosol-free atmosphere is also given. (left) An integration angle of $\zeta = 1.5^\circ$ is used -- typical value used at the Pierre Auger Observatory. (right) An integration angle of $\zeta = 3.0^\circ$ is used. Colours in online version.}
\label{fig:ms_vs_elevangle}
\end{figure*}

It is also interesting to study the dependence of the indirect light fraction on the elevation angle of the fluorescence detector. Indeed, an observation at lower inclination angles would be more affected by multiple scattering since the part of atmosphere probed by the pixels is much more dense in scatterers (molecules and aerosols). Figure~\ref{fig:ms_vs_elevangle} gives the indirect light fraction versus the inclination angle of the detector for the same vertical air shower, for different values of asymmetry parameter, and for $\zeta=1.5^\circ$ (left) or $\zeta = 3.0^\circ$ (right). As it was expected, this indirect light contribution is higher in the lower part of the atmosphere. As already observed previously in the same section, larger aerosols (i.e.\ higher values for the asymmetry parameter $g$) produce a greater effect on the indirect light recorded by the detector. Such plots guide us to identify the parameters for which the multiply scattered fraction is dependent: the distance to the air shower, the aerosol conditions (aerosol attenuation length and aerosol size) and the integration angle on the camera $\zeta$. The goal of the next section is to define properly this parameterisation depending on all these parameters.

\section{Parameterisation of the multiply scattered fraction}
\label{sec:parameterisation}

\begin{figure*}[p]
\centering
\includegraphics [width=1.0\textwidth] {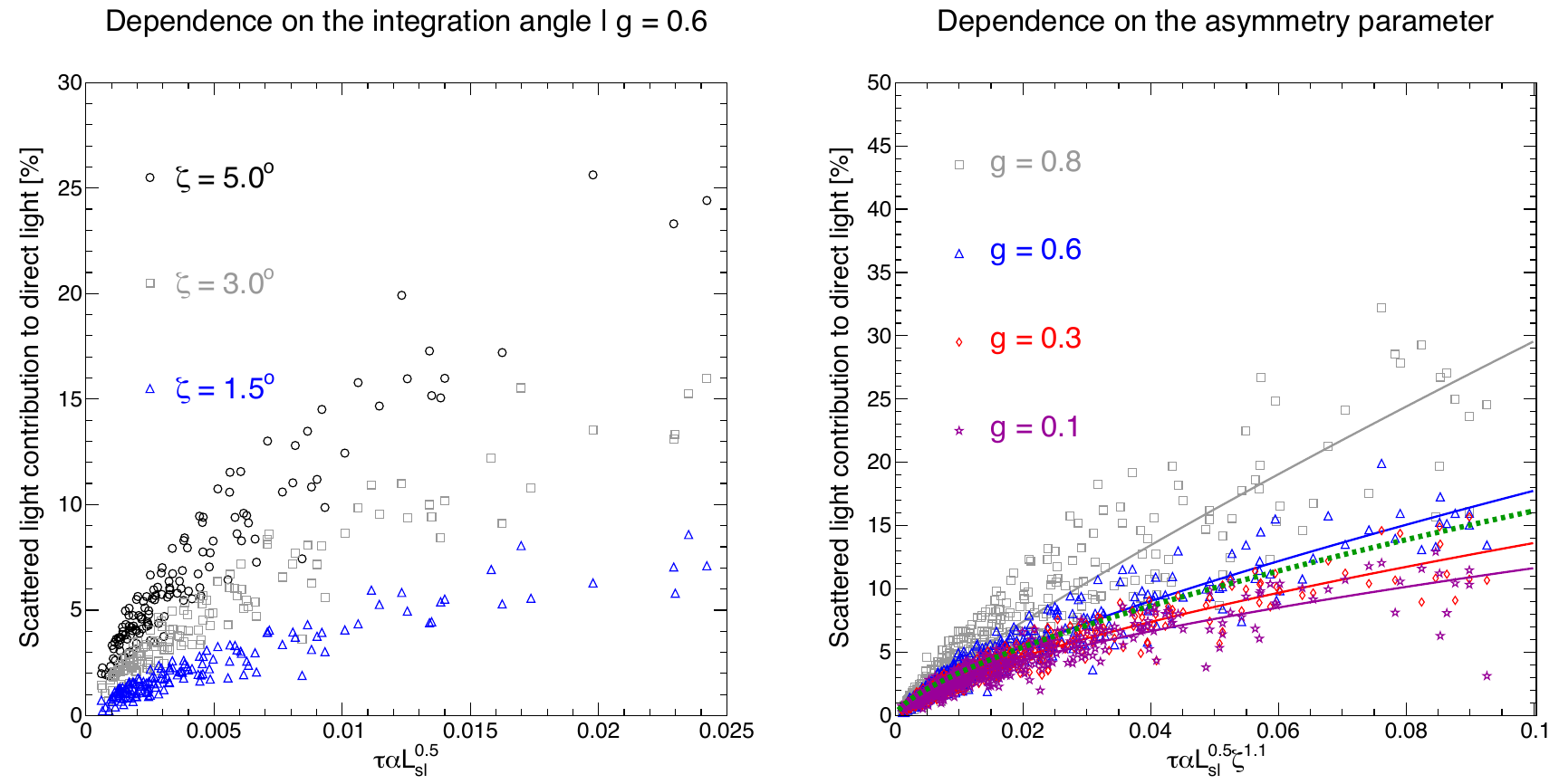}
\caption{{\bf Parameterisation of the multiple scattered fraction and dependence on geometry of air showers, light collection on camera and aerosol conditions.} (left) Indirect light contribution from Monte Carlo simulations $\kappa_{\rm MC}$ plotted versus a combination of geometric and atmospheric parameters. Each marker represents the indirect light contribution estimated with one Monte Carlo simulation. A clear correlation is observed with this combination $\tau \alpha L^{0.5}$ and a strong dependence on the integration angle $\zeta$ is noticed. Only simulations with an asymmetry parameter of $g=0.6$ are plotted here. (right) Same plot but including the integration angle $\zeta$ in the combination of parameters. $\kappa_{\rm MC}$ values (i.e.\ markers) for other values of asymmetry parameter $g$ are also plotted. The continuous lines $\kappa_{\rm PAR}$ correspond to the fit of $\kappa_{\rm MC}$ values using Eq.~\eqref{eq:param_model} for each population of $g$ values. The dotted line corresponds to the parameterisation found by M.\ D.\ Roberts in~\cite{MS_roberts}. Colours in online version.}
\label{fig:ms_parameterisation}
\end{figure*}

After having studied in detail the properties of indirect light arriving at a ground-based detector, the purpose of this section is to find a global parametric model for the multiply scattered fraction. A large library of Monte Carlo simulations was produced, varying the geometric and atmospheric parameters. Only vertical air showers were simulated as explained in Sect.~\ref{sec:simu_results}. Different distances between the detector and the core of the air shower $L$ were simulated: $5~$km, $10~$km, $20~$km, $30~$km and $40~$km. Also, different viewing angles $\theta_{\rm inc}$ for the fluorescence detectors are probed: $3^\circ$, $6^\circ$, $9^\circ$, $12^\circ$, $15^\circ$, $18^\circ$, $21^\circ$, $24^\circ$, $27^\circ$ and $30^\circ$. For each of these geometric configurations, different aerosol conditions were also simulated with values of aerosol attenuation length $\Lambda^0_{\rm{aer}}$ of $10~$km, $25~$km and $50~$km, and values of asymmetry parameter $g$ of $0.1$, $0.3$, $0.6$ and $0.8$. The vertical aerosol scale $H^0_{\rm{aer}}$ is fixed at $1.5~$km for the whole set of simulations. Values of these parameters were chosen to probe a wide range of light source positions and aerosol scattering conditions.

\begin{figure}[!t]
\centering
\vspace{-0.8cm}
\includegraphics [width=0.50\textwidth] {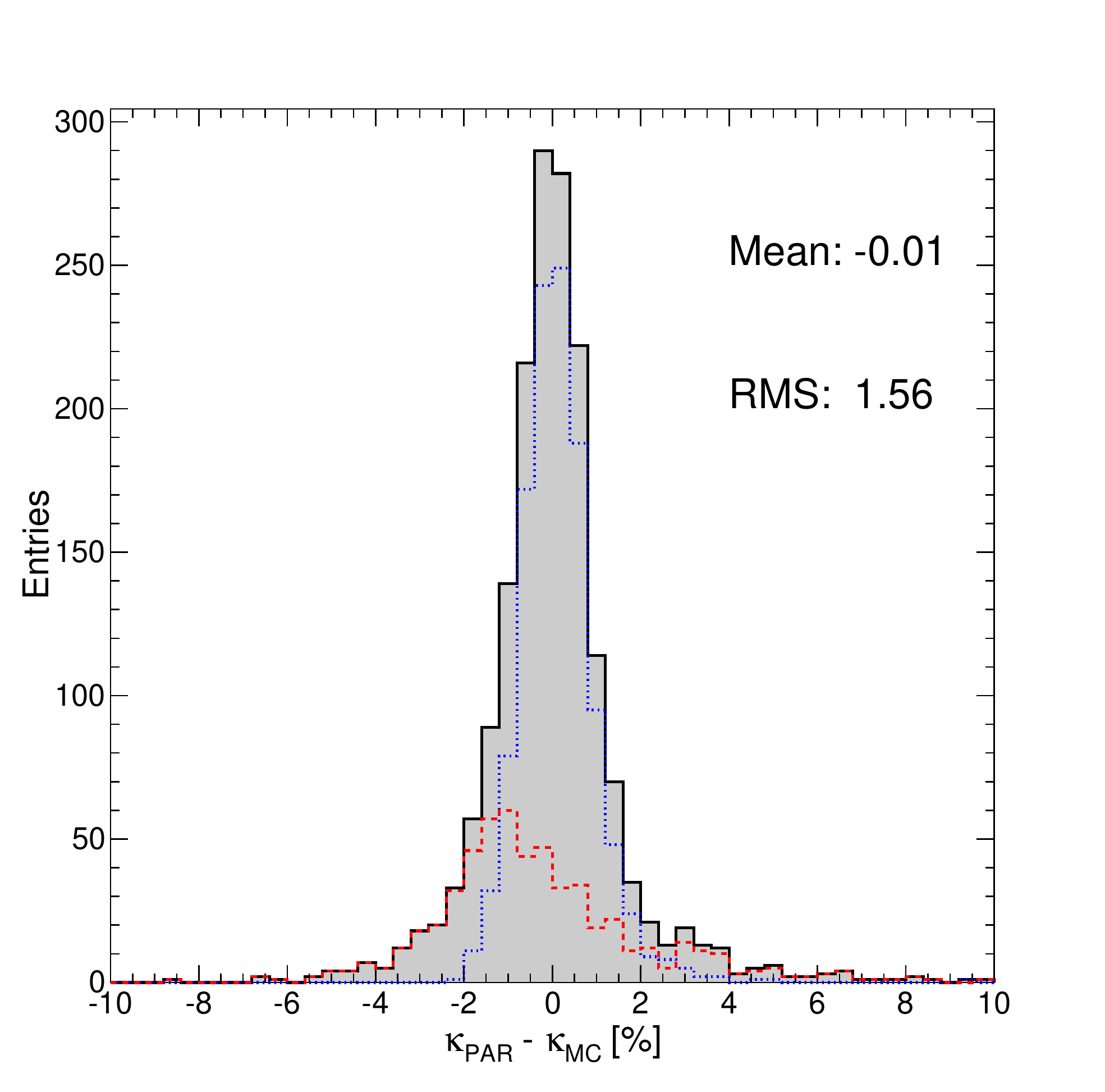}
\caption{{\bf Distribution of residuals between parameterised values $\kappa_{\rm PAR}$ and simulated values $\kappa_{\rm MC}$ for the indirect light contribution.} The same distributions are also plotted for simulations for $\kappa_{\rm MC}$ values lower than $5\%$ {\it (dotted line)} and greater than $5\%$ {\it (dashed line)}. Colours in online version.}
\label{fig:ms_residuals}
\end{figure}

New parameters are now defined to make simpler the work. For instance, instead of using the elevation of the direct light source $\theta_{\rm inc}$ and the aerosol and molecular attenuation lengths $\Lambda^0_{\rm{aer}}$ and $\Lambda^0_{\rm{mol}}$, the total optical depth $\tau$ is used. $\tau$ is equal to $-\ln(N^{\rm det}_{\rm dir}/N)$, where $N$ is the total number of photons generated in the solid angle of the detector, and $N^{\rm det}_{\rm dir}$ the number of direct photons recorded by the detector. The parameter $\tau$ represents the total attenuation corresponding to molecular and aerosol components between two points in the atmosphere. The total scattering coefficient at the position of photon production $\alpha$ is also useful in the following. It is defined as $1/\Lambda_{\rm{aer}}(h_{\rm agl}) + 1/\Lambda_{\rm{mol}}(h_{\rm agl})$, where $h_{\rm agl}$ is the altitude of the photon source above ground level. Values of multiply scattered fraction $\kappa_{\rm MC}$ estimated using the Monte Carlo simulation library are then used to develop the global parameterisation to predict the multiply scattered fraction $\kappa_{\rm PAR}$. Based on the parametric model already developed by M.\ D.\ Roberts in~\cite{MS_roberts} for extensive air showers, the parameters are combined with power law dependences to get the best agreement between $\kappa_{\rm MC}$ and $\kappa_{\rm PAR}$. Figure~\ref{fig:ms_parameterisation} (left) plots the $\kappa_{\rm MC}$ values for all Monte Carlo simulations done with an asymmetry parameter $g$ of $0.6$. Multiply scattered fractions are given for three different integration angles $\zeta$, from $1.5^\circ$ to $5.0^\circ$. A clear trend is observed for these $\kappa_{\rm MC}$ values when they are plotted as a function of the quantity $\tau \alpha L^{0.5}_{\rm sl}$, where $L_{\rm sl}$ is the slant length between the detector and the source, i.e.\ $L_{\rm sl} = L/\cos \theta_{\rm inc}$. However, we observe three distinct populations corresponding to the three different integration angles $\zeta$. Then, the $\zeta$ dependence is also included in the combination of parameters. $\kappa_{\rm MC}$ values against this new combination are plotted in Fig.~\ref{fig:ms_parameterisation}~(right) for the same $g$ value of $0.6$, but also for other values of asymmetry parameter. Finally, a two-parameter power law fit to $\kappa_{\rm MC}$ values for each population of asymmetry parameter gives the final form for the parametric model
\begin{equation}
\kappa_{\rm PAR} (g) [\%] = A(g)\times \left(\tau \alpha L^{0.5}_{\rm sl} \zeta^{1.1}\right)^{B(g)},  
\label{eq:param_model}
\end{equation}
where $\{A(g),\,B(g) \}$ are two fit parameters defined as $A(g)$ $= 596\times g^5 + 52$ and $B(g) = 0.35\times g + 0.56$. The parameter $L_{\rm sl}$ is given in meters, the total scattering coefficient $\alpha$ in m$^{-1}$ and the integration angle $\zeta$ in degrees. The parametric models $\kappa_{\rm PAR}$ for each $g$ value are plotted in continuous lines in Fig.~\ref{fig:ms_parameterisation}~(right). This parameterisation depending on the asymmetry parameter is also compared to the original M.\ D.\ Roberts' parametric model given in dotted line. Since the aerosol phase function used in this original work was similar to a DHG function with a $g=0.6$, it is reasonable to find this model close to our $\kappa_{\rm PAR}$ curve representing this same $g$ value. In order to check the validity of this parametric model, the residuals between $\kappa_{\rm PAR}$ and $\kappa_{\rm MC}$ are plotted in Fig.~\ref{fig:ms_residuals} for all the configurations probed in the library of Monte Carlo simulations. No systematic bias is observed in the residuals. However, a smooth tail is observed, increasing the RMS of the distribution. To better understand this tail, the distribution is divided into two populations corresponding to $\kappa_{\rm MC}$ lower (dotted line) or greater (dashed line) than $5\%$. The result obtained is convincing since residuals increasing this tail correspond to large indirect light fractions. Deviations of the parametric model to simulated values are similar to other works done in~\cite{MS_pekala,MS_roberts}. Also, this dispersion and the induced error is lower than uncertainties that we have in the determination of the geometric parameters of the air shower and the aerosol conditions in a UHECR experiment. The purpose of the last section is to estimate the systematic effect of the multiple scattering in the atmosphere on the energy and $X_{\rm max}$ reconstruction for extensive air showers.

\section{Air shower reconstruction and systematic effect on energy and $X_{\rm max}$}
\label{sec:eas}

\begin{figure*}[!t]
\centering
\includegraphics [width=1.0\textwidth] {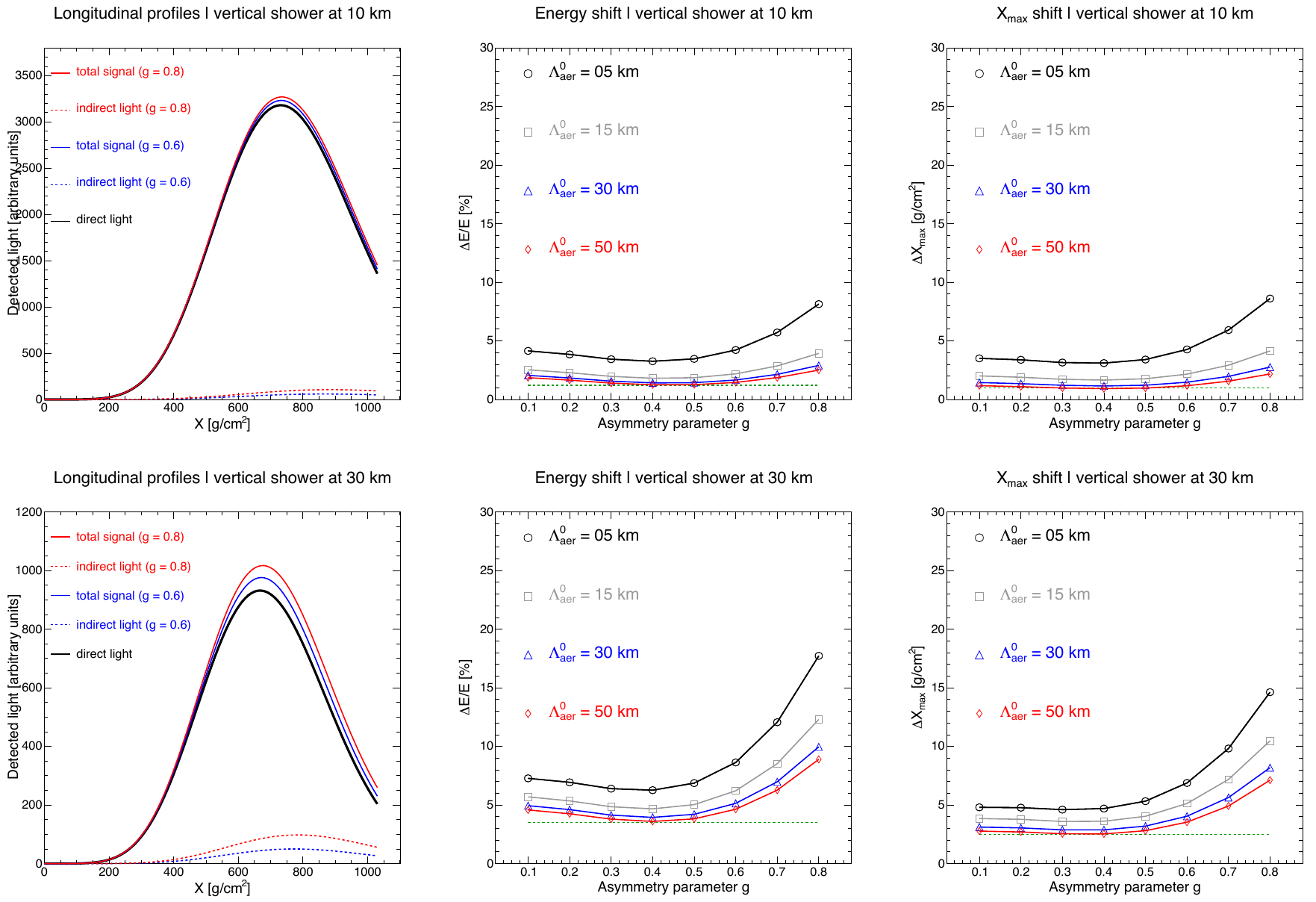}
\caption{{\bf Reconstruction of the longitudinal profile for a vertical air shower located at 10~km and 30~km.} (left) The thick line represents the longitudinal profile composed only of direct photons collected by the detector. Indirect photons are given in dashed lines for an asymmetry parameter $g$ of $0.6$ or $0.8$. The profiles giving the total signal are also plotted for the two cases. The aerosol attenuation length $\Lambda^0_{\rm{aer}}$ is fixed at $25~$km and the integration angle $\zeta=1.5^\circ$. (middle) Energy shift induced by the indirect light for different aerosol attenuation lengths and values of asymmetry parameter. (right) $X_{\rm max}$ shift induced by the indirect light for different aerosol attenuation lengths and values of asymmetry parameter. The dotted line corresponds to the case of an aerosol-free atmosphere. Colours in online version.}
\label{fig:ms_airshowers}
\end{figure*}

The air-fluorescence technique, combined to a surface detector composed of water-Cherenkov tanks or scintillators, has the advantage to set the energy scale of an "hybrid" observatory. Since the fluorescence light produced is directly proportional to the energy of the UHECR inducing the extensive air shower, the fluorescence detector provides a nearly calorimetric energy measurement. Also, the depth of maximum of the extensive air shower, called $X_{\rm max}$, gives a direct way to constrain the UHECR composition. In the reconstruction of these two quantities, atmospheric effects on propagation of fluorescence photons have to be taken into account carefully. For instance, for an air shower located at $30~$km from the fluorescence telescope, it is more than $60\%$ of the total light which is attenuated (this estimation is obviously an average value since it is well-known that aerosol population fluctuates strongly in time). Thus, it is important to well estimate the indirect light contribution to the total signal recorded by the detector. Indeed, neglecting the multiple scattered light contribution to total signal would over-estimate systematically the energy and the $X_{\rm max}$ of the extensive air showers. This effect is currently taken into account in UHECR ground-based experiments using parameterisations developed in~\cite{MS_giller,MS_roberts,MS_pekala} to avoid this systematic over-estimation. In this part, a typical value of $1.5^\circ$ is used for the integration angle on the fluorescence telescope. Figure~\ref{fig:ms_airshowers}~(left) gives the longitudinal profiles for the same vertical air shower located at $10~$km or $30~$km from the detector. For both geometries, the multiply scattered light contribution is given alone or added to the direct light, as a function of the slant depth $X$. As expected, this contribution is stronger at higher values of slant depth, i.e.\ in the lower part of the atmosphere. Comparing the profiles corresponding to the direct light with the total one, it is easy to observe the distortion in the shower development profile induced by the multiply scattered component. Figure~\ref{fig:ms_airshowers}~(middle) gives the energy shift and Figure~\ref{fig:ms_airshowers}~(right) the $X_{\rm max}$ shift induced by the multiple scattering, for different values of aerosol attenuation length and asymmetry parameter. As already known with the previous works done by M.\ D.\ Roberts~\cite{MS_roberts}, J.\ Pekala~\cite{MS_pekala} or M.\ Giller~\cite{MS_giller}, the energy and $X_{\rm max}$ shifts increase with optical depth. However, none of them have studied the dependence of these shifts on the asymmetry parameter (i.e.\ the aerosol size): this work presented in the paper quantifies finally these two shifts and shows that they are clearly dependent on the asymmetry parameter too. In the case of a distant ($30~$km) extensive air shower with an aerosol attenuation length $\Lambda^0_{\rm{aer}}$ greater than $15~$km, multiple scattering leads to a systematic over-estimation of $5.0\pm1.5\%$ [$5.0^{\rm + 7.0}_{\rm - 1.5}\%$] for the energy and $4.0\pm 1.5~$g/cm$^2$ [$4.0^{\rm + 6.5}_{\rm - 1.5}~$g/cm$^2$] for the $X_{\rm max}$ in the case of $g$ values lower than $0.7$ [$0.8$], where the uncertainties refer to the variation of the aerosol size. This error is especially high if aerosols of large size are present in the atmosphere. The different plateaus observed at low $g$ values result from a combination of different effects: the number of fluorescence photons scattered close to their production point, their attenuation during their propagation to the detector and the asymmetry parameter.  

After the understanding and the observation of the dependence of the multiply scattered fraction on the asymmetry parameter, it is natural to think how it could affect UHECR measurements for the ground-based observatories. Large atmospheric monitoring programs were developed at the UHECR observatories, especially to characterise the aerosol component. Among these properties, the aerosol scattering phase function was directly measured by the HiRes~\cite{APF_hires} and the Pierre Auger~\cite{APF_pao,MyICRC} collaborations. They measured a value for the asymmetry parameter of about $0.6\pm0.1$. Contrary to the optical depth measurements, the asymmetry parameter is not measured every night and only an average value of $0.6$ is used in the air shower reconstruction. Thus, using this new parameterisation including a dependence on the asymmetry parameter, a systematic uncertainty induced by the aerosol size on the energy and $X_{\rm max}$ could be estimated for each UHECR observatory. This value would depend on the characteristics of the site and on the quality cuts placed on the data in the science analysis. Referring to previous measurements of the asymmetry parameter $g$ and to Fig.~\ref{fig:ms_airshowers}, their systematic uncertainties associated to the aerosol size would be only a few percents for energy and a few g/cm$^2$ for $X_{\rm max}$.

\section{Conclusion}
A Monte Carlo simulation for the scattering of light was adapted to study the effect of aerosols on the multiple scattering in the air shower reconstruction. The contribution of indirect photons to the total fluorescence light induced by an air shower and recorded by a ground-based detector was investigated in detail. An overview of the different properties of indirect photons was presented, allowing us to understand better how aerosols and especially their size can affect the multiple scattered fraction recorded by the detector.

A complete parameterisation of the multiple scattered fraction, including for the first time the dependence on the asymmetry parameter of the aerosol scattering phase function, was determined. This parameterisation can predict the indirect light fraction for any geometries of air shower and any aerosol conditions, and could be easily included in a program of air shower reconstruction. For a vertical extensive air shower observed during typical aerosol conditions by a ground-based detector at $30~$km, multiple scattering leads to a systematic over-estimation of $5\pm1.5\%$ for the energy and $4.0\pm 1.5~$g/cm$^2$ for the $X_{\rm max}$, where the uncertainties refer to a variation of the asymmetry parameter ($0.1 \leqslant g \leqslant 0.7$).

\section*{Acknowledgements}
K.\ L.\ thanks Marcel Urban for having provided the stimulus to begin this study. Also, the authors thank their colleagues from the Pierre Auger Collaboration and the referees for fruitful discussions and/or for their comments on this work. Further gratitude is expressed to the team of technicians maintaining the computer cluster CC IN2P3 where simulations were run.



\end{document}